\documentstyle[aps,prl]{revtex}
\begin{document}
\author{Y.M. Vilk, and A.-M.S. Tremblay}
\title{Theory of single-particle properties of the Hubbard model}
\date{24 January 1995}
\address{D\'epartement de Physique and Centre de Recherche en Physique du
Solide,\\
Universit\'e de Sherbrooke, Sherbrooke, Qu\'ebec, Canada J1K 2R1}
\maketitle

\begin{abstract}
It is shown that it is possible to quantitatively explain quantum Monte
Carlo results for the Green's function of the two-dimensional Hubbard model
in the weak to intermediate coupling regime. The analytic approach includes
vertex corrections in a paramagnon-like self-energy. All parameters are
determined self-consistently. This approach clearly shows that in two
dimensions Fermi-liquid quasiparticles disappear in the paramagnetic state
when the antiferromagnetic correlation length becomes larger than the
electronic thermal de Broglie wavelength.
\end{abstract}


For more than thirty years, the Hubbard model has served as a theoretical
basis for a wide range of interacting electron phenomena, from itinerant
electron magnetism to, more recently, high-temperature superconductivity. In
the latter context, one of the key issues has been the possibility of
non-Fermi liquid behavior in the single-particle properties. These
properties have been intensively studied in the weak to intermediate
coupling regime using paramagnon-type theories.\cite{Schrieffer}\cite{Kampf}
Such approaches are physically attractive since they describe the effect of
low-lying collective modes on single-particle properties in a manner similar
to that of phonons. However in these theories one faces the difficulty that,
contrary to the case of phonons, vertex corrections are not small (no Migdal
theorem) and have to be taken into account.

In this letter, we introduce a simple approach that has the straightforward
physical interpretation of paramagnon theories and uses no adjustable
parameter. It does include vertex corrections, and achieves, for both
two-particle and single-particle properties, better quantitative agreement
with Monte Carlo simulations than previous theories\cite{FLEX}\cite{parquet}%
\cite{FLEX-parquet}. Our analytical approach allows us to unambiguously
interpret the Monte Carlo data and to show that in two dimensions spin
fluctuations destroy the Fermi liquid in the paramagnetic state when the
antiferromagnetic correlation length becomes larger than the thermal de
Broglie wavelength of electrons. This corresponds to the disappearance of
the quasiparticle peak and the appearance of a pseudogap, an issue that has
been controversial mostly due to finite-size effects in Monte Carlo\cite
{White}.

We consider the one-band Hubbard model on the square lattice with unit
lattice spacing, on-site repulsion $U$ and nearest-neighbor hopping $t$. We
work in units where the lattice spacing is unity, $k_B=1$, $\hbar =1$ and $%
t=1$. The theory has a simple structure that we now explain physically,
postponing the formal derivation momentarily.

The calculation proceeds in two steps: we first obtain spin and charge
susceptibilities, then we inject them in the self-energy calculation. In the
calculation of susceptibilities we make the approximation that spin and
charge susceptibilities $\chi _{sp}$, $\chi _{ch}$ are given by RPA-like
forms but with two different effective interactions $U_{sp}$ and $U_{ch}$
which are then determined self-consistently. The necessity to have two
different effective interactions for spin and for charge is dictated by the
Pauli exclusion principle $\langle n_\sigma ^2\rangle =\langle n_\sigma
\rangle $ which implies that both $\chi _{sp}$ and $\chi _{ch}$ are related
to only one local pair correlation function $\langle n_{\uparrow
}n_{\downarrow }\rangle $. Indeed, using the Fluctuation-Dissipation theorem
in Matsubara formalism and the Pauli principle one can write:

\begin{equation}
\frac 1{\beta N}\sum_q\chi _{ch,sp}(q)=\frac 1{\beta N}\sum_q\frac{\chi _0(q)%
}{1+\frac{\left( -1\right) ^\ell }2U_{ch,sp}\chi _0(q)}=n+2\left( -1\right)
^\ell \langle n_{\uparrow }n_{\downarrow }\rangle -\left( 1-\ell \right) n^2
\label{sumSpin}
\end{equation}
where $\ell =0$ for charge $\left( ch\right) $, $\ell =1$ for spin $(sp)$, $%
\beta \equiv 1/T$, $n=\langle n_{\uparrow }\rangle +\langle n_{\downarrow
}\rangle $, $q=({\bf q},iq_n)$ with ${\bf q}$ the wave vectors of an $N$
site lattice, $iq_n$ Matsubara frequencies and $\chi _0(q)$ the
susceptibility for non-interacting electrons. The value of $\langle
n_{\uparrow }n_{\downarrow }\rangle $ may be obtained self-consistently\cite
{Vilk1} by adding to the above set of equations the relation $%
U_{sp}=g_{\uparrow \downarrow }(0)\,U$ with $g_{\uparrow \downarrow
}(0)\equiv \langle n_{\uparrow }n_{\downarrow }\rangle /\langle
n_{\downarrow }\rangle \langle n_{\uparrow }\rangle .$

As shown in Ref.\cite{Vilk1}, the above procedure reproduces both
Kanamori-Brueckner screening as well as the effect of Mermin-Wagner thermal
fluctuations, giving a phase transition only at zero-temperature in
two-dimensions. There is however a crossover temperature $T_X$ below which
the magnetic correlation length $\xi $ grows exponentially. Quantitative
agreement with Monte Carlo simulations is obtained\cite{Vilk1} for all
fillings and temperatures in the weak to intermediate coupling regime $U<8$.

We now turn to the discussion of the single-particle properties. In order,
to be consistent with the two-particle correlation functions, the
self-energy $\Sigma _\sigma \left( k\right) $ must satisfy the sum rule
\begin{equation}  \label{sumS}
\lim _{\tau \rightarrow 0^{-}}\frac 1{\beta N}\sum_k\Sigma _\sigma \left(
k\right) G_\sigma \left( k\right) e^{-ik_n\tau }=U\left\langle n_{\uparrow
}n_{\downarrow }\right\rangle ,
\end{equation}
which follows from the definition of $\Sigma _\sigma \left( k\right) $ \cite
{BaymKadanoff}. Here, we encounter the same key quantity $\langle
n_{\uparrow }n_{\downarrow }\rangle $ that appears in the sum rule for the
susceptibilities Eq. (\ref{sumSpin}). We find, using the formal approach
described below, the following expression for $\Sigma _\sigma \left(
k\right) $

\begin{equation}  \label{param}
\Sigma _\sigma \left( k\right) =Un_{-\sigma }+\frac U4\frac TN\sum_q\left[
U_{sp}\chi _{sp}(q)+U_{ch}\chi _{ch}(q)\right] G_\sigma ^0(k+q),
\end{equation}
which satisfies Eq. (\ref{sumS}) with $G_\sigma $ replaced by $G_\sigma ^0$
on the left-hand side. This self-energy expression (\ref{param}) is
physically appealing since, as expected from general skeleton diagrams, one
of the vertices is the bare one $U$, while the other vertex is dressed and
given by $U_{sp}$ or $U_{ch}$ depending on the type of fluctuation being
exchanged. Eq.(\ref{param}) already gives good agreement with Monte Carlo
data but the accuracy can be improved even further by requiring that the
consistency condition (\ref{sumS}) be satisfied with $G_\sigma $ instead of $%
G_\sigma ^0$. To do so we replace $U_{sp}$ and $U_{ch}$ on the right-hand
side of (\ref{param}) by $\alpha U_{sp}$ and $\alpha U_{ch}$ with $\alpha $
determined self-consistently by Eq.(\ref{sumS}). For $U<4$, we have $\alpha
<1.15$. This concludes the description of the structure of our theory. We
now turn to the formal derivation.

{\em Formal aspects: }We use the notation $1=\left( {\bf R}_1,\tau _1\right)
$, and an overbar $\left( \overline{1}\right) $ to indicate a space-time or
spin variable which is integrated or summed over. Using the Matsubara
Green's function $G_\sigma \left( 1,3\right) $ in the presence of an
external potential $\phi _\sigma \left( 2\right) $, one can write the exact
equations for the generalized susceptibilities $\chi _{_{\sigma \sigma
^{\prime }}}\left( 1,3;2\right) \equiv \delta G_\sigma \left( 1,3\right)
/\delta \phi _{\sigma ^{\prime }}\left( 2\right) $ and the self-energy $%
\Sigma _\sigma \left( 1,2\right) $ as\cite{BaymKadanoff}:

\begin{equation}  \label{dGdPhi}
\chi _{\sigma ,\sigma ^{\prime }}\left( 1,3;2\right) =G_\sigma \left(
1,2\right) G_\sigma \left( 2,3\right) \delta _{\sigma ,\sigma ^{\prime
}}+G_\sigma \left( 1,\overline{6}\right) G_\sigma \left( \overline{7}%
,3\right) \Gamma _{\sigma ,\bar \sigma }^{ir}\left( \overline{6},\overline{7}%
;\overline{4},\overline{5}\right) \chi _{\overline{\sigma },\sigma ^{\prime
}}\left( \overline{4},\overline{5};2\right)
\end{equation}

\begin{equation}
\Sigma _\sigma \left( 1,2\right) =Un_{-\sigma }\delta \left( 1-2\right)
-UG_\sigma \left( 1,\overline{6}\right) \Gamma _{\sigma \bar \sigma
}^{ir}\left( \overline{6},2;\overline{4},\overline{5}\right) \chi _{_{\bar
\sigma ,-\sigma }}\left( \bar 4,\bar 5;1\right) .  \label{SigmaG}
\end{equation}
where $\Gamma _{\sigma ,\sigma ^{\prime }}^{ir}=\delta \Sigma _\sigma
/\delta G_{\sigma ^{\prime }}$ are irreducible vertices. Our approximation
for spin and charge susceptibilities and for self-energy Eq.(\ref{param}) is
obtained from these exact equations by using as initial guess on the
right-hand side of {\em both} equations a self-energy $\Sigma _\sigma ^{(0)}$
and irreducible vertices $\Gamma _{\sigma \bar \sigma }^{(0)ir}$ which are
functional derivatives of

\begin{equation}  \label{Phi}
\Phi \left[ G\right] =\frac 14(U_{ch}-U_{sp})\widetilde{n}_{\overline{\sigma
}}\left( \overline{1}\right) \widetilde{n}_{\overline{\sigma }}\left(
\overline{1}\right) +\frac 14(U_{ch}+U_{sp})\widetilde{n}_{\overline{\sigma }%
}\left( \overline{1}\right) \widetilde{n}_{-\overline{\sigma }}(\overline{1})
\end{equation}
where $\widetilde{n}_\sigma \left( 1\right) \equiv G_\sigma \left(
1,1^{+}\right) $. The initial self-energy $\Sigma _\sigma ^{(0)}=\delta \Phi
/\delta G_\sigma |_{\phi =0}$ is a constant and hence can be absorbed in a
chemical potential shift so that all Green's functions on the right-hand
side of Eqs. (\ref{dGdPhi}) and (\ref{SigmaG}) are bare ones $G_\sigma
^0\left( 1,2\right) $. The corresponding particle-hole irreducible vertices $%
\Gamma ^{\left( 0\right) ir}=\delta ^2\Phi /(\delta G^2)|_{\phi =0}$ are
given in Fourier space by two constants
\begin{equation}  \label{defVertex}
\Gamma _{\sigma ;-\sigma }^{(0)ir}=(U_{ch}+U_{sp})/2\quad ;\quad \Gamma
_{\sigma ,\sigma }^{(0)ir}=(U_{ch}-U_{sp})/2
\end{equation}
which are chosen so that the susceptibilities are consistent with the
correct double-occupancy $\langle n_{\uparrow }n_{\downarrow }\rangle $ and
with the exclusion principle $\langle n_\sigma ^2\rangle =\langle n_\sigma
\rangle $, as described above.

Important advantages of our iterative procedure are that: a) The
approximation is conserving since $\Sigma _\sigma ^{(0)}$ and $\Gamma
_{\sigma ,\sigma ^{\prime }}^{\left( 0\right) ir}$ are functional
derivatives of a single functional $\Phi $\cite{Baym}. b) The following
identity
\begin{equation}
\Sigma _\sigma \left( 1,\overline{3}\right) G_\sigma \left( \overline{3}%
,2\right) =-U\left\langle T_\tau \left[ n_{-\sigma }\left( 1\right) \psi
_\sigma \left( 1\right) \psi _\sigma ^{+}\left( 2\right) \right]
\right\rangle   \label{consistency}
\end{equation}
is satisfied with $\Sigma _\sigma ^{\left( 1\right) }$ and $G_\sigma
^{\left( 0\right) }$ and it can be used as a convergence check, namely $%
\Sigma _\sigma ^{\left( 1\right) }\left( 1,\overline{3}\right) G_\sigma
^{\left( 1\right) }\left( \overline{3},1^{+}\right) \approx \Sigma _\sigma
^{\left( 1\right) }\left( 1,\overline{3}\right) G_\sigma ^{\left( 0\right)
}\left( \overline{3},1^{+}\right) $. For $U<4$, this convergence criterion
is satisfied at worse within $15\%$. The sum rule (\ref{sumS}) follows from
the identity (\ref{consistency}) when $2=1^{+}$.

{\em Comparisons with other theories and with Quantum Monte Carlo data:}
Fig.1(a) shows $G\left( {\bf k,}\tau \right) $ for filling $n=0.875$,
temperature $T=0.25$ and $U=4$ for the wave vector on the $8\times 8$
lattice which is closest to the Fermi surface, namely $\left( \pi ,0\right) $%
. For these parameters, size effects are negligible. Our theory is in
agreement with Monte Carlo data and with the parquet approach but in this
regime second-order perturbation theory for the self-energy gives the same
result. This surprising performance of perturbation theory (see also \cite
{Pertub}) is a consequence of compensation between the renormalized vertices
and susceptibilities ($U_{sp}<U$, $\chi _{sp}(q)>\chi _0(q)$; $U_{ch}>U$, $%
\chi _{ch}(q)<\chi _0(q)$ ).

Half-filling $n=1$ is an ideal situation for numerical studies of low-energy
phenomena since some of the allowed wave vectors on finite lattices lie
exactly on the Fermi surface. Fig.1(b) shows $G({\bf k}_F,\tau )$ for ${\bf k%
}_F{\bf =}\left( \pi ,0\right) $ in a regime where the antiferromagnetic
correlation length is growing exponentially. The Green's function is
strongly renormalized since the non-interacting value is $-0.5$, independent
of $\tau $. Our theory shows better agreement with Monte-Carlo than previous
approaches. Despite the very large correlation length, finite-size effects
for the Matsubara Green's function $G({\bf k}_F,\tau )$ are not large and
almost completely disappear when the system size becomes larger than the
thermal de Broglie wavelength of the electrons ($\xi _{th}\equiv
\left\langle v_F\right\rangle /\pi T$) as we explain below.

Our most dramatic numerical results addressing the issue of the Fermi liquid
are shown in Fig. 2 where we plot
\begin{equation}  \label{z}
\tilde z\left( T\right) =-2G\left( {\bf k}_F,\beta /2\right) =\int \frac{%
d\omega }{2\pi }\frac{A\left( {\bf k}_F,\omega \right) }{\cosh \left( \beta
\omega /2\right) }.
\end{equation}
This quantity $\tilde z\left( T\right) $ is an average of the
single-particle spectral weight $A\left( {\bf k}_F,\omega \right) $ within $%
T\equiv 1/\beta $ around the Fermi level ($\omega =0$) and it is a
generalization of the usual zero-temperature quasiparticle renormalization
factor $z\equiv 1/(1-\partial \Sigma /\partial \omega )$. For
non-interacting particles $\tilde z\left( T\right) $ is unity. For a normal
Fermi liquid it becomes equal to a constant less than unity as the
temperature decreases since the width of the quasiparticle peak scales as $%
T^2$ and hence lies within $T$ of the Fermi level. The quantity $\tilde z%
\left( T\right) $ is the best estimate of $z$ one can get from Monte Carlo
data for $G({\bf k},\tau )$. Moreover $\tilde z\left( T\right) $ gives an
estimate of $A\left( {\bf k}_F,\omega \right) $ around the Fermi surface
even when the Fermi liquid does not exist.

One can clearly see from Fig. 2 that while second-order perturbation theory
exhibits typical Fermi-liquid behavior for $\tilde z\left( T\right) $, both
Monte Carlo data and a numerical evaluation of our expression for the
self-energy lead to a rapid fall-off of $\tilde z\left( T\right) $ below $%
T_X $ (for $U=4$, $T_X\approx 0.2$\cite{Vilk1}). The decrease of $\tilde z%
\left( T\right) $ in this regime is approximately $\tilde z\sim T^2$ which
clearly suggests non-Fermi liquid behavior. The physical origin of this
effect is that the quasi-particles of the {\em two-dimensional }paramagnetic
state become overdamped when the energy scale associated with the proximity
to antiferromagnetism $\delta U\equiv U_{mf,c}-U_{sp}$ ($U_{mf,c}\equiv
2/\chi _0\left( {\bf Q},0\right) $) becomes exponentially small or, more
precisely, when the two-particle antiferromagnetic correlation length $\xi $
becomes larger than the single-particle thermal de Broglie wavelength $\xi
_{th}\equiv v_F/\pi T$.

{\em Pseudogap: }While size effects and statistical errors make continuation
of the Monte Carlo data to real frequencies particularly difficult, in our
approach we can make this continuation analytically to show that the above
effect corresponds to a pseudogap. We first explain the behavior $\tilde z%
\sim T^2$ then do real frequency analysis.

Since the spin susceptibility $\chi _{sp}\left( {\bf q},0\right) $ below $%
T_X $ is almost singular at the antiferromagnetic wave vector ${\bf Q}%
=\left( \pi ,\pi \right) $, the main contribution to $\Sigma $ in Eq. (\ref
{param}) comes from $iq_n=0$ and wave vectors $({\bf q-Q)}^2\leq \xi ^{-2}$
near ${\bf Q}$. Approximating $\chi _{sp}\left( {\bf q},0\right) $ in Eqs. (%
\ref{sumSpin}) and (\ref{param}) by its asymptotic form $\chi _{sp}\left(
{\bf q},0\right) \approx 2\left[ U_{sp}\xi _0^2(\xi ^{-2}+({\bf q-Q)}%
^2)\right] ^{-1}$ where $\xi _0^2\equiv \frac{-1}{2\chi _0\left( Q\right) }%
\frac{\partial ^2\chi _0\left( Q\right) }{\partial q_x^2}$ and $\xi \equiv
\xi _0(U_{sp}/\delta U)^{1/2}$, the integrals over ${\bf q}$ can be done to
obtain the asymptotic results

\begin{equation}  \label{ksi}
\xi \sim \exp \left( \pi \tilde \sigma ^2\xi _0^2\frac{U_{sp}}T\right)
\end{equation}

\begin{equation}  \label{SigmaM}
\Sigma \left( {\bf k}_F,ik_n\right) =\frac U2-i\frac{UT}{8\pi \xi _0^2\sqrt{%
k_n^2-v_F^2\xi ^{-2}}}\ln \frac{k_n+\sqrt{k_n^2-v_F^2\xi ^{-2}}}{k_n-\sqrt{%
k_n^2-v_F^2\xi ^{-2}}}+{\cal R}.
\end{equation}
Here ${\cal R}$ is the regular part which remains finite as $T\rightarrow 0$
and $\tilde \sigma ^2\equiv n-2\langle n_{\uparrow }n_{\downarrow }\rangle
-C<1$ is the right-hand side of (\ref{sumSpin}) minus corrections $C$ that
come from the sum over non-zero Matsubara frequencies (quantum effects) and
from $({\bf q-Q)}^2\gg \xi ^{-2}$. The corresponding value of $\tilde z(T)\ $%
can be written as the alternating series $-2G\left( {\bf k}_F,\beta
/2\right) =-4T\sum_{n=1}^\infty \left( -1\right) ^n/\left( ik_n+\mu -\Sigma
\left( {\bf k}_F,ik_n\right) \right) $. We can understand qualitatively both
the temperature and size dependence of the Monte Carlo data in Fig. 2 by
using the first term of this series along with Eqs. (\ref{ksi}) and (\ref
{SigmaM})
\begin{equation}
\tilde z(T)\sim \frac{T^2}{\tilde \sigma ^2UU_{sp}}\sqrt{1-\frac{\xi _{th}^2%
}{\xi ^2}}.
\end{equation}
On the infinite lattice, $\xi $ starts growing exponentially below $T_X$,
quickly becoming much larger than $\xi _{th}$. This implies $\tilde z%
(T)\simeq T^2$. On finite lattices Eq.(\ref{sumSpin}) gives $\xi \sim \sqrt{N%
}$ below $T_X$ which explains the size effect observed in Monte Carlo {\it %
i.e.} smaller $\tilde z$ for smaller size $N$, ($\xi _{th}(T_X)\sim 5$ for
Fig. 2).

The analytic continuation of $\Sigma \left( {\bf k}_F,ik_n\right) $ in Eq. (%
\ref{SigmaM}) is

\begin{equation}
\Sigma ^R\left( {\bf k}_F,\omega \right) =\frac U2+\frac{UT}{8\pi \xi _0^2%
\sqrt{\omega ^2+v_F^2\xi ^{-2}}}\ln \left| \frac{\omega +\sqrt{\omega
^2+v_F^2\xi ^{-2}}}{\omega -\sqrt{\omega ^2+v_F^2\xi ^{-2}}}\right| -i\frac{%
UT}{8\xi _0^2\sqrt{\omega ^2+v_F^2\xi ^{-2}}}+{\cal R.}  \label{Somega}
\end{equation}
At small frequencies the regular part ${\cal R}$ can be neglected so that
exactly at the Fermi surface $\left( \omega =0\right) $ the imaginary part
of the self-energy for $\xi >\xi _{th}$ increases exponentially when the
temperature decreases, $\Sigma ^{\prime \prime }({\bf k}_F,\omega )\sim U\xi
/(\xi _{th}\xi _0^2)\sim \exp \left( \pi \tilde \sigma ^2\xi
_0^2U_{sp}/T\right) $\cite{footnote1}. This corresponds to a pseudogap:
Instead of a quasiparticle peak, the spectral weight has a minimum at the
Fermi level and two symmetrically located maxima away from it. Clearly this
is not a Fermi liquid, yet the symmetry of the system remains unbroken at
any finite $T$.  By contrast\cite{footnote2} with $3D$, in $2D$ this effect
exists in a wide temperature range $T<T_X$. These conclusions persist
slightly away from half-filling. In particular, we do not find a
quasiparticle peak in the pseudogap close to half-filling when $\xi \gg \xi
_{th}$. This is different from the results inferred from a phenomenological
zero-temperature calculation\cite{Kampf} $\left( \xi _{th}=\infty \right) $
which physically corresponds to $1\ll \xi \ll \xi _{th}$.

In photoemission experiments on real quasi two-dimensional materials we
predict a rapid decrease of the spectral weight and density of states at the
Fermi level in a wide temperature range, from $T_X$ to the N\'eel
temperature $T_N$ ($T_X-T_N\sim 10^2\,\,K$).

We thank Liang Chen, R. C\^ot\'e and A.E. Ruckenstein for numerous
enlightening discussions. We acknowledge financial support from the Natural
Sciences and Engineering Research Council of Canada (NSERC), the Fonds pour
la formation de Chercheurs et l'aide \`a la Recherche from the Government of
Qu\'ebec (FCAR), and (A.-M.S.T) the Canadian Institute of Advanced Research
(CIAR) and the Killam foundation.

\figure{Fig. 1. Comparison of our results for $G({\bf k},\tau )$ with Monte
Carlo data, FLEX, parquet, and second-order pertubation theory, all on $%
8\times 8$ mesh. Monte Carlo data and results for FLEX and parquet are from
Ref.\cite{FLEX-parquet}. a) $n=0.875$, $T=0.25$; b) $n=1$, $T=0.17$.}

\figure{Fig. 2. Temperature dependence of the generalized renormalization
factor $\tilde z$ defined in Eq.(\ref{z}). Lines are results of our
calculations for infinite lattice and $16\times 16$ mesh. Symbols are Monte
Carlo data from Ref. \cite{White}.}


\begin{references}
\bibitem{footnote1}  It can be shown that for $T>T_X$, $\xi \sim 1$ the
regular part ${\cal R}$ dominates in equation (\ref{Somega}) and one
recovers standard behavior with the maximum in the spectral weight at the
Fermi level.
\end{references}
\end{document}